\documentstyle[prl,twocolumn,aps,psfig]{revtex}

\begin{document}
\title{Front dynamics in turbulent media  }
\author{A.C. Mart\'\i ,$^{1}$ 
F. Sagu\'es,$^{2}$ and J.M. Sancho$^{1}$}

\address{
$^1$ Departament d'Estructura i Constituents de la Mat\`eria,\\
Universitat de Barcelona,\\
Av. Diagonal 647,\\
E-08028 Barcelona, Spain.\\
$^2$ Departament de Qu\'{\i}mica F\'{\i}sica,\\
Universitat de Barcelona,\\
Av. Diagonal 647,\\
E-08028 Barcelona, Spain.\\
}
\date{\today}
\maketitle

\begin{abstract} 
A study of a stable front propagating in a turbulent medium is presented.
The front is generated through a reaction-diffusion equation, and the
turbulent medium is statistically modeled using a Langevin equation.
Numerical simulations indicate the presence of two different dynamical
regimes. These regimes appear when the turbulent flow either wrinkles a
still rather sharp propagating interfase or broadens it. Specific
dependences of the propagating velocities on stirring intensities
appropriate to each case are found and fitted when possible according to
theoretically predicted laws. Different turbulent spectra are considered. 
\end{abstract}


\section{Introduction}
\label{sec:intr}

Front propagation has been a problem of great interest in a rich variety
of nonequilibrium phenomena originating pattern structures
\cite{pelce,murray,cross}. The most usual and studied situation
corresponds to a stable (planar) front propagating in a quiescent
homogeneous and isotropic medium at a constant velocity.  This situation
can be modeled easily by a reaction-diffusion equation for the order
parameter. Reaction (nonlinear) terms have to present at least two
possible steady states: one stable and a second unstable or metastable. 
Then the stable state propagates through the unstable one. The region
where the variable changes abruptly from one state to the other is called
the interface or front width, whose size is controlled by the diffusion
coefficient. These fronts propagate at a velocity $v_0$ which results from
the interplay between the chemical time scale (reaction) $\tau _{\rm chem}$
and the species diffusivity coefficient (diffusion) $D$: $v_0\sim
(D\tau^{-1} _{\rm chem})^{1/2}$.  These facts are well known and have been
discussed and applied in different fields. 

A more complicated situation is found when considering front--like
patterns propagating in non--quiescent media. In such a case, the
additional length and time scales introduced by the advecting flow
will interact with those intrinsic to the reaction-diffusion front
dynamics, resulting in new and distinctive propagation
regimes. Particularly interesting in this context is the problem of
reaction propagation in premixed combustible gases \cite{Williams85}.
Mainly due to its practical relevance in combustion processes, the
subject has motivated numerous theoretical as well as experimental
efforts
\cite{Pope87,Ronney95a,Kernstein88,Kernstein92,Majda96,Pocheau}. The
generic understanding coming from these studies is that the turbulent
front velocity $v_T$ is larger than $v_0$, to an extent which depends
on the intensity and spatiotemporal correlations of the turbulent
flow. Some other issues, however, such as velocity quenching effects,
role of turbulent spectra remain somewhat more open.

An enlarged perspective has been gained recently when experiments on
liquid phase reactions \cite{Shy92} have brought out a more
simplified scenario as compared to that of combustion processes.  In
such liquid reactions the change of density and the increasing of the
temperature are negligible compared with those occurring in typical
combustion processes \cite{Ronney95}.  In this way experimentalists
are closer to reproduce the set of simplified assumptions invoked by
most of the theoretical models.

An even more controlled situation concerning liquid phase reactions
can be envisaged when considering front propagation under externally
imposed stirring conditions. This is precisely the context we will
address in this paper.  In essence, stirring can be thought of as a
sort of controlled way to inject energy into a fluid medium through
random forces of statistical nature
\cite{Batchelor70,McComb90and95}. Especially suited to reproduce such
an scenario are the generically called \cite{Juneja94} synthetic
turbulence generating models. In particular we will employ here an
algorithm to generate stochastic, turbulent-like, flows satisfying
Langevin type equations \cite{marti}. Specifically, random flows with
zero mean velocity and statistically isotropic, homogeneous and
stationary are created independently of the front conditions.

The closest experimental reference we are aware concerning this
question is the recently reported experimental work by Ronney et
al. \cite{Ronney95} dealing with different stirred media. From the
theoretical side however there is no specific literature focusing on
stirring effects, so we rely on the general references of front
propagation under turbulent flows.  Most of the existing theoretical
models refer to the so-called Huygens mechanism valid for thin front
conditions (``flamelet regime''). Among them we mention those based on
the so-called G-equation \cite{Kernstein88}, nonlinear averaging of
reaction-diffusion equations \cite{Majda96}, and specially, for the
purpose of comparison with our results here, the approaches of
Kernstein \cite{Kernstein92}, Yakhot \cite{Yakhot88} and Pocheau
\cite{Pocheau}.  These authors based on different assumptions proposed
scaling relations for $v_T$ in terms of the intensity of the flow
$u_0^2$. Using the so--called flame propagation equation, a variant of
the well-known KPZ-equation, Kernstein and Ashurst \cite{Kernstein92}
obtained a relation valid for weak stirring intensities and large
correlation times of the flow. Applying renormalization procedures to
the G-equation, Yakhot \cite{Yakhot88} found a relation without
adjustable parameters. Pocheau claimed that both Kernstein and Yakhot
laws converge into a quadratic relation when invariance over all the
scales is assumed \cite{Pocheau}.

On the other hand, when relying on computer simulations \cite{Haworth92},
numerical accuracy and finite size effects commonly
restricts the examined scenarios to small turbulent intensities.  It is
also worth mentioning here the alternative stochastic method reviewed by
Pope \cite{Pope87}. In such a formulation, a probability density functional
is introduced and, after invoking some closure approximation and
elimination of variables procedure, a tractable set of partial
differential equations is obtained.  The most important difficulties of
this method remain at the fundamental level of the closure approximation. 

In comparison with these generic procedures, what we gain with our
random flow generating algorithm is an easy control over the most
representative statistical quantities of the advecting flow: its
energy spectrum and spatiotemporal correlations. This in turn enables
us to explore the role of these parameters in determining the
different propagation regimes and particularly in the enhancement of
the front speed under turbulent conditions.  In particular, the two
basic experimentally identified modes of front propagation under
stirring \cite{Ronney95}, i.e. the previously mentioned Huygens
propagation mode and the distributed reaction zone regime, are
qualitatively identified and quantitatively analyzed in our
simulations. Moreover we are able to explore for each regime the role
of a pair of different energy spectra displaying a quite different
range of energy containing modes.

The paper is organized as follows. In the next section we summarize
the method to simulate the turbulent medium. In Sec. \ref{sec:results}
we introduce the theoretical model of front propagation. Numerical
results are presented and commented according the existing theoretical
predictions. We devote Sec. \ref{sec:summ} to draw some conclusions
and perspectives.

\section{Turbulent media modelization}
\label{sec:lang}

Here we will present a brief summary of the theoretical scheme to
simulate a statistically homogeneous, stationary and isotropic
turbulent flow characterized by its energy spectrum. As we want to
study fronts dynamics under different turbulent conditions we will
generate two types of spectra.

As a first example we adopt the Kraichnan's (K) spectrum
\cite{Kraichnan70}, describing a distributed band of excitations
around a well pronounced peak centered at some well-defined wave
number $k_0$,
\begin{equation}
\label{Kraichnan}
E(k)\propto k^3\exp \left[ -{\frac{3 k^2}{{2 k_0^2}}}\right].
\end{equation}

In order to take into account the influence of a broader spectrum of
modes we will modelize another type of turbulent medium with the
choice of the K\'arman-Obukhov's (KO) spectrum which was introduced
\cite{K-O} to study Kolgomorov turbulence.  To this end we select from
the family of KO spectra, the following form:
\begin{equation}
E(k) \propto k^3 \left[ 1 + {9 k^2 \over{5 k_0^2}} \right] ^{ -7/3}
\label{eq:K-O}
\end{equation}
which has its maximum also at $k_0$.

As we will assume that the characteristics of the flow are not going
to be affected by front dynamics, then our method will simply
reproduce the above mentioned spectra in a statistical way (see
Ref.~\cite{marti} for more theoretical details).

Our starting point is a generalized Ornstein-Uhlenbeck like Langevin
equation for the stream function $\eta ({\bf r},t)$,
\begin{equation}
\label{langevin}{\frac{\partial \eta ({\bf r},t)}{{\partial t}}}=\nu \nabla
^2\eta ({\bf r},t)%
+Q[\lambda ^2\nabla ^2]\nabla .\mbox{\boldmath $ \zeta $}({\bf r},t)
\end{equation}
where $\nu$ is the kinematic viscosity and 
$\mbox{\boldmath $ \zeta $}({\bf r},t)$ is a Gaussian white noise 
process of zero mean and correlation
\begin{equation}
\label{ruido}\left\langle \zeta ^i({\bf r}_1,t_1)\zeta ^j({\bf r}%
_2,t_2)\right\rangle =2\epsilon _0\nu \delta (t_1-t_2)\delta ({\bf r}_1-{\bf %
r}_2)\delta ^{ij}.
\end{equation}

According to these definitions, $\epsilon_0$ and $\lambda $ are
control parameters respectively related to the intensity and
characteristic length of the random flow.  $ Q[\lambda ^2\nabla^2]$
plays a relevant role in our approach because it introduces through
the random stirring forces the spectra we want to use in each
particular case. In particular for the K spectrum we take
\begin{equation}
Q[\lambda ^2\nabla^2]
=\exp \left( {\frac{\lambda ^2\nabla ^2}{{2}}}\right)
\end{equation}
where $ \lambda = (3/2)^{1/2}k_0^{-1}$. 
On the other hand, the KO spectrum
is reproduced through
\begin{equation}
Q[ \lambda^2 \nabla^2] =  \left( 1 - \lambda^2 \nabla^2  \right)^{-7/6}
\label{eq:QforK-O}
\end{equation}
where $ \lambda = (9/5)^{1/2}k_0^{-1}$. 

From the stream function, the 2-d incompressible advecting flow is
obtained as usual,
\begin{equation}
\label{defvel}
{\bf v }({\bf r},t) = \left( -{\frac{{\partial \eta ({\bf r},t)} }{{%
\partial y}}},{\frac{{\ \partial \eta ({\bf r},t)} }{{\ \partial x }}}\right) .
\end{equation}

This two dimensional version of our algorithm is chosen for the
simplicity in dealing with the numerical simulations, but at the same
time expresses our belief that the basic trends of front propagation
in turbulent media, that we want to analyze here, are well reproduced
in this reduced dimensionality.  Moreover, recent experiments in
chemical reactions in quasi 2-d geometries \cite{Ronney95} and
theoretical analysis \cite{Pocheau} support also this assumption.

The energy spectrum $E(k)$ can be expressed in terms of the Fourier
transformed operator $Q(- \lambda^2 k^2)$,
\begin{equation}
E(k) = { \epsilon_0 \over {4\pi}}k^3 Q^2 (- \lambda^2 k^2).
\label{eq:ek}
\end{equation}

The stirring intensity, $u_0^2$, ($u_0^2= \int_0^{\infty}{dk E(k)} $) and
integral time $t_0$ and length $l_0$ scales of the random flows are easily
expressed in terms of the functional and parametric specifications for the
noise terms left in (\ref{langevin}) and (\ref{ruido}). 

In particular for the K spectrum, we have
\begin{equation} 
u_0^2=\frac{\epsilon _0}{8\pi \lambda ^4};  \,\,\,\,\,
t_0=\frac{\lambda ^2}\nu ; \,\,\,\,\,
l_0=\frac{\lambda \sqrt{\pi }}2. 
\end{equation}
And for KO spectrum, we get
for the three basic parameters
\begin{equation}
u_0^2 = \frac{9 \epsilon_0}{32 \pi \lambda^4};   \,\,\,\,\, 
t_0 =  \frac{\lambda^2}{3 \nu};
\end{equation}
\begin{equation}
l_0 = \lambda {\Gamma(1/2) \Gamma(5/6) \over{2 \Gamma(1/3)}}.
\end{equation}

In Fig.~\ref{fig1} three different spectra with the same intensity
(kinetic energy) are presented: one of Kraichnan type (a) and two of
the K\'arman-Obukhov spectrum b) and c). We see how Kraichnan' s
energy distribution shows a pronounced maximum whereas the other
spectra are broader and show much longer tails. Spectra (a) and (c)
have the same $k_0$, but (a) and (b) have the same $l_0$. These facts
will be of relevance later on to explain the dynamical properties of a
front propagating in random media represented by each one of those
spectra.

For the sake of comparison, we have considered two additional and
somewhat related, turbulent conditions. The first one, hereafter
referred on as frozen stirring, corresponds to a frozen configuration
of the random flow generated according the rules above introduced. The
second one referred on what follows as periodic flow, represent
nothing but a periodic set of fixed eddies.  It is constructed from
the single mode stream function
\begin{equation}
\eta (x,y)=\eta_0 
\cos ( \frac{n \pi x}{N}) \cos(\frac{n\pi y}{N}),
\end{equation}
representing a periodic array of $n \times n$ eddies, where $N$ is
defined from the system size $L$ as, $N=L/ \Delta x $ .

A detailed presentation of the way the algorithm just proposed is
implemented to simulate turbulent flows can be found in \cite{marti}.


\section{Front dynamics}
\label{sec:results}

\subsection{The model and theoretical basis}

In this section we present the model used to reproduce a chemical
stable front propagating in random media with prescribed statistical
properties.

The basic ingredient is a reaction-diffusion scheme which exhibits two
steady states of different stability and connected by a front which
otherwise propagates in a stable (planar) way. There are many possible
choices in the literature \cite{murray,cross} and we will take a very
simple one which corresponds to that of a generic reaction-diffusion
equation for the scalar field $\psi ({\bf r},t)$, used as a relevant
variable:
\begin{equation}
{\frac{\partial \psi }{{\partial t}}}=D\nabla ^2\psi + f(\psi)
\label{eq:psi}
\end{equation}
where the nonlinear function  $f(\psi)$ has a minimum of two zeros 
corresponding to the steady states.
In our numerical simulations we have chosen the nonlinear function 
$f(\psi)=\psi^2 - \psi^3$.
In this case a stable planar front propagates the stable state $\psi =1$
(``products'') into the invaded metastable one $\psi =0$ (``reactants''). 
Two important parameters are the dimensionless propagation rate and front
thickness which respectively read: 
\begin{equation}
v_0=\sqrt{\frac{D}{2
}},\,\,\,\,\,\,\,\, \delta_0= \sqrt{ 8 D}.
\label{eq:quo}
\end{equation}
The front profile has then the form,
\begin{equation}
\psi( x, t) = \frac{1}{2} 
\left( 1 - \tanh( \frac{x-v_0 t}{\delta_0}) \right).
\end{equation}
Our numerical results have been checked in relation with different choices of
the reaction term. Confirming our belief that within our formulation, this 
choice should not be a relevant point, we have explicitly considered the case 
 $f(\psi) = \psi(1-\psi)(\psi - 0.25)$ and the
same quantitative results have been obtained. 

A turbulent flow is superposed now through a convective term added to
Eq.~(\ref{eq:psi}) 
\begin{equation}
{\frac{\partial \psi }{{\partial t}}}=D\nabla ^2\psi +\psi ^2-\psi ^3
 - {\bf \nabla}.[{\bf v}({\bf r},t) \psi].
\label{eq:psi-con}
\end{equation}
As illustrated in Figs. \ref{fig2}a and \ref{fig2}b, numerical
simulations of Eq.~(\ref{eq:psi-con}) above, reproduce two distinct
limiting regimes of front propagation. In these two figures the flow
pattern is also plotted. Small arrows indicate the local velocity
vector of the flow.

In Fig. \ref{fig2}a we see a front in a situation where the typical
length scale of the flow $l_0$ is larger than the intrinsic one
associated to the reaction-diffusion dynamics $\delta _0$. It is clear
that the mean size of the observed eddies is larger than the front
width. Then a distorted front propagates with a larger velocity than
in the deterministic case but maintaining a still rather sharp and
well defined interface. Such a propagation mechanism is known in the
combustion literature as the ``thin flame,'' ``flamelet'' or
``reaction sheet'' regime \cite{Ronney95}.

The other limit corresponds to a situation for which $l_0$ is smaller than
$\delta _0$. Now the mean size of the eddies are comparable or smaller
than the interfacial width. This is the situation in Fig. \ref{fig2}b
where the front is broader and more diffused than in the former case. This
situation is referred in the literature as a ``distributed reaction
zone''(DRZ) regime \cite{Ronney95}. Also in this case we also observe that
the front velocity is larger than in quiescent media. 

We want to emphasize at this point that both phenomenologies are well
known in experiments dealing with front propagation for isothermal
chemical reactions \cite{Ronney95}. 

A specific mechanism of front propagation applies to each one of the
previously identified modes. The common rationale behind the ``thin
flame'' mode is based on a HP-like argument: the front has the same
local structure as in the planar case with normal velocity given by
$v_0$, but its length increases due to wrinkling and this results on
faster propagation velocities. This can be understood as a geometrical
consequence of the propagation of curved interface with a local
velocity $v_0$ \cite{Kernstein88}. Denoting respectively by $L_T$ and
$L_0$ the front length and the lateral system size, we have
\begin{equation}
\frac{L_T}{L_0}=\frac{v_T}{v_0}\equiv S.
 \label{HP:eq}
\end{equation}

On the other hand if we assume that the effect of the flow velocity in the
DRZ regime is completely reproduced by increasing the diffusive transport
inside the broadened front then such an effect can be incorporated as a
renormalization of the diffusion coefficient. In the DRZ mode we simply
adapt the first fundamental result of Eq.~\ref{eq:quo}, to obtain

\begin{equation}
S^2 =  \left( \frac{v_T}{v_0} \right)^2 = \left( \frac{D_T}D\right)
\label{DRZ:eq}
\end{equation}
where $D_T$ is the effective turbulent diffusion.
The next and most involved step consists, however, in using 
Eqs.~(\ref{HP:eq}) and
(\ref{DRZ:eq}) above to make
detailed predictions for $v_T$, or its
dimensionless form $S$, as a function of the stirring intensity $u_0$, or its
dimensionless value $u_0 /v_0 \equiv Q$. Let us discuss on
what follows our numerical results for the HP and DRZ regimes.

\subsection{Results for HP versus DRZ propagation modes}

In Figs.~(\ref{fig3})--(\ref{fig5}) we present our numerical results
corresponding to these two modes of front propagation. If it is not
explicitly indicated the turbulent flow is that of K spectrum.  On
what refers to the HP mode, our first task was to check relation
(\ref{HP:eq}). The collected data for the different values of $u_0^2$
are summarized in Fig.~\ref{fig3}. For the sake of comparison, we
include in this figure results obtained for the other two additional
stirring conditions introduced in Sec.  \ref{sec:lang}: the frozen
stirring and the periodic flow. According to this figure, the
geometric argument leading to (\ref{HP:eq}) seems well-supported by
our simulations with the simple exception of those situations
involving very intense periodic flows. Actually, under these last
conditions the interface is largely perturbed by the presence of
overhangs whose dynamics contribute positively to the computed
velocity, measured as the time variation of the rate of occupation of
the $\psi =1$ state, but negatively to the front length.

In Fig.~\ref{fig4}, results for $S(Q)$ are specifically plotted
corresponding to three of the previously mentioned stirring modes.
Figure~\ref{fig4}a shows the simulation results of $S^2$ versus $Q^2$
for a turbulent flow for two different values of $t_0$.  The
theoretical predictions of Yakhot \cite{Yakhot88}, ($ S = \exp( Q^2/
S^2)$) and Pocheau ($S^2 = 1 + 2Q^2$) \cite{Pocheau} are also
plotted. Numerical data of the case of frozen stirring are included
for comparison.  Actually, the numerical results fit reasonably well
the linear relation $S^2=1+ \alpha Q^2$, for the whole range of
$Q$. Nevertheless, the slope of such a linear law depends on $t_0$, a
fact not considered in theoretical predictions of Yakhot and
Pocheau. Remarkably, these theoretical predictions fit better with our
numerical results for frozen stirring (limit of very large
$t_0$). This agreement is not surprising because in Pocheau's
analysis, both theoretical and experimental, a very large value of
$t_0$ was considered \cite{Pocheau}. Moreover, for this last flow, the
behavior largely depends on the examined range of $Q$: the former
behavior transforms into a $ S=1+ \alpha ' Q^{\frac 43}$ when
approaching the smallest values of $Q$ here considered
(Fig.~\ref{fig4}b).

Finally, results for the periodic stirring also show a crossover from
a linear dependence at large $Q$ towards the quadratic form: $ S^2=1+
\alpha'' Q^2$ at small $Q$ (Fig.~\ref{fig4}c). Results for the last
two cases are in agreement with theoretical predictions by Kernstein
et al.  \cite{Kernstein92}.

A final comment is worth emphasizing at this point: Turbulent
propagation velocities fall always lower than those obtained for
frozen and periodic stirring. This is somewhat at odds with what is
reported in Refs~\cite{Ronney95a,Ronney95}, although one should keep
in mind that the range of $Q$ values there considered were of two
orders of magnitude larger than ours.

In Fig.~\ref{fig5} results for the DRZ regimen are presented. The
numerical data obtained under different stirring mechanisms are
plotted.  These numerical results are compared with the corresponding
theoretically predicted values based on Eq.~(\ref{DRZ:eq}) above.  The
values of the effective diffusion coefficient $D_T$ have been obtained
independently from direct simulation of pure scalar diffusion (without
reaction).  Theoretical predictions based in Eq.~(\ref{DRZ:eq})
exhibit a remarkable agreement with numerical results for a broad set
of the $Q$ values here considered and irrespective of the type of
stirring flow considered.  Moreover, the theoretical dependences of
$D_T$ on $u_0^2$ either for random flows in the weak stirring limit
($D_T -D \sim u_0^2 t_0$) \cite{marti} or for periodic flows in the
limit of small Peclet number ($D_T -D \sim u_0^2 /D$) \cite{Moffat},
are clearly confirmed in this Fig.~\ref{fig5}.

\subsection{Results for K versus KO spectra}

\begin{figure} 
\centerline{\psfig{figure=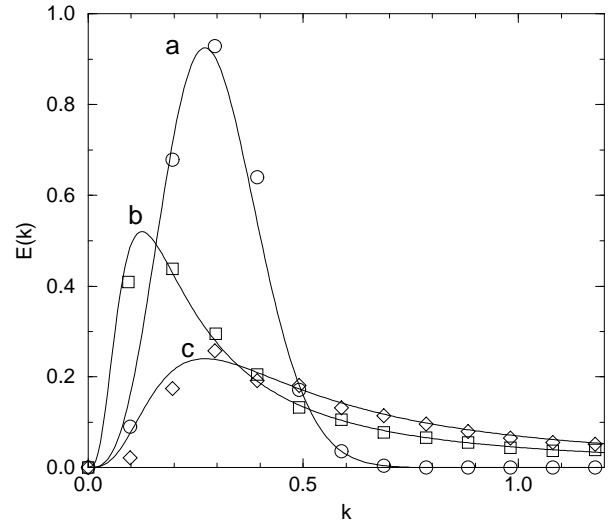,height=7cm}}
\caption{Comparison of Kraichnan (a) and K\'arman--Obukhov (b) and
(c) spectra.  Continuous lines correspond to the analytical expression 
Eq.~(\protect\ref{eq:ek}) and symbols stand for
simulation results in the discrete
lattice. (a) Kraichnan, $l_0=4.0$, $k_0=0.27$; (b) and (c) 
K\'arman-Obukhov $l_0=4.0$,
$k_0=0.12$ and $l_0=1.85$, $k_0=0.27$ respectively.
The intensity (kinetic energy) is fixed to $u_0^2=0.25$. 
Unless otherwise stated a square lattice of $128$x$128$ 
points and unit spacing $\Delta=0.5$ has been employed in all our 
simulations.}
\label{fig1}
\end{figure}

\begin{figure} 
\centerline{\psfig{figure=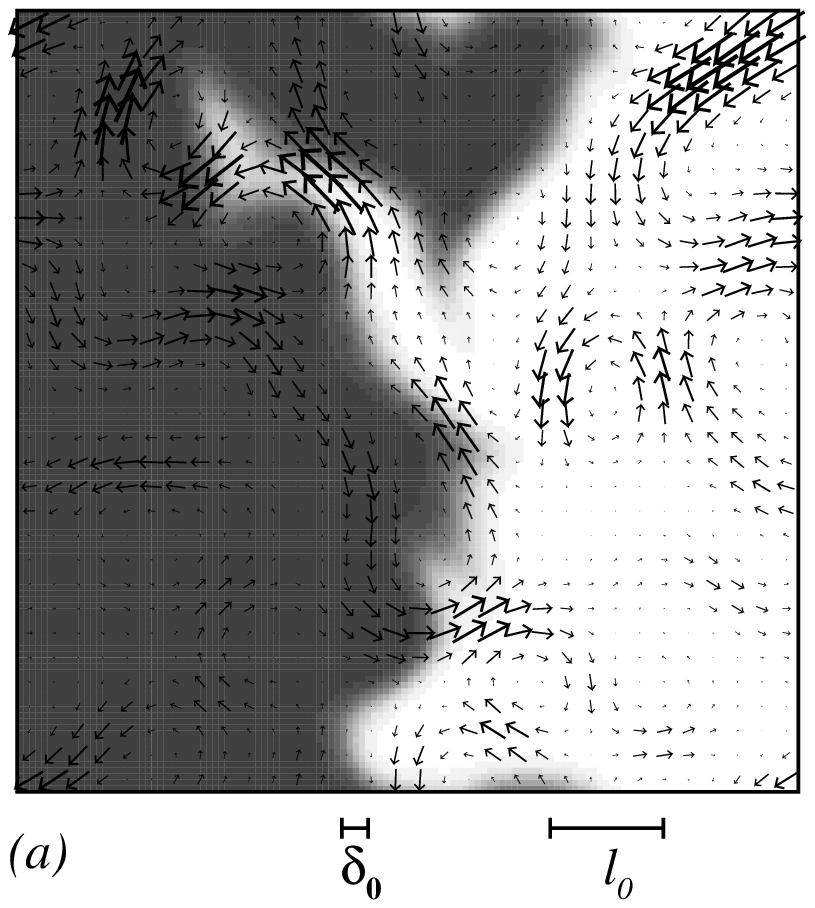,height=7cm}}
\centerline{\psfig{figure=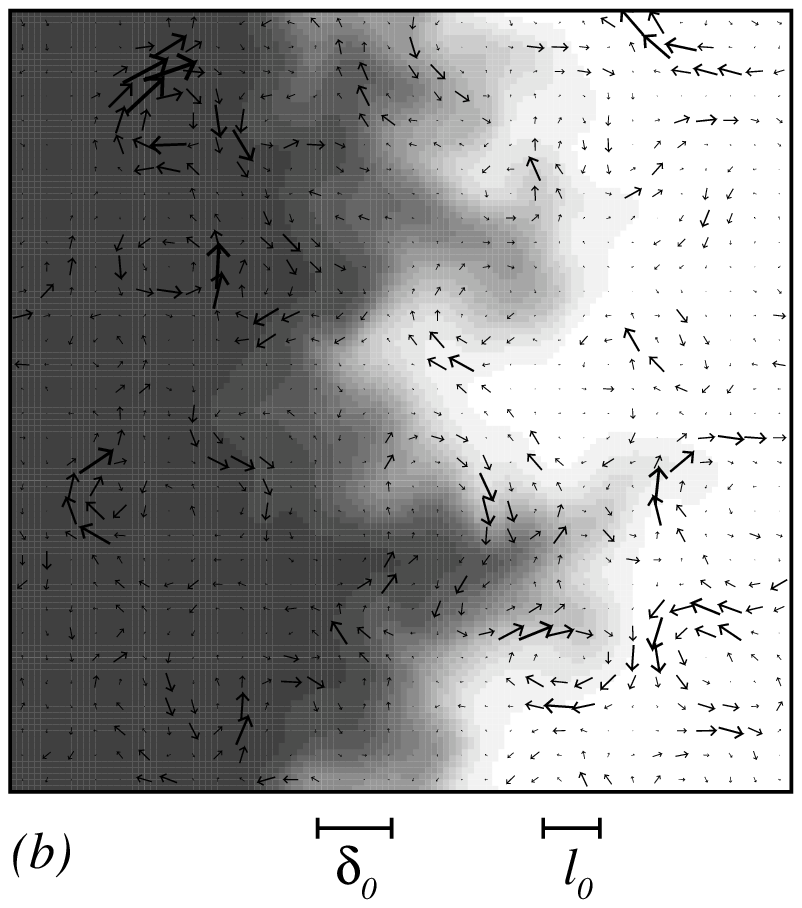,height=7cm}}
\caption{Density plot of the stirred reaction front for the K spectrum.
The arrows stand for the local velocity field plotted every 
$4$ mesh points in each direction. (a) corresponds to the thin front mode, 
parameter values are
$D=0.3$, $u_0^2=3.0$, $\lambda=4.5$ and $t_0=1.0$ (b) corresponds to the
distributed reaction front regime for $D=2.0$, $u_0^2=10.0$,
$\lambda=2.2$, $t_0=1.0$.} 
\label{fig2} 
\end{figure}

\begin{figure} 
\centerline{\psfig{figure=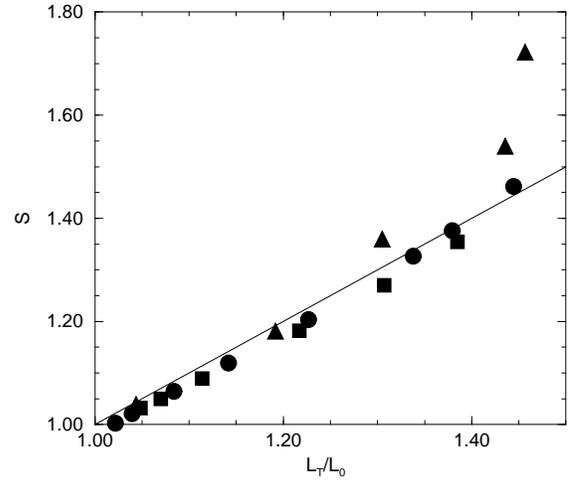,height=7cm}}
\caption{ $S$ vs $L_T / L_0$ for turbulent flow
(circles, $l_0=4.0$, $t_0=0.1$), random frozen (squares, $l_0=4.0$) and
periodic eddies (triangles, $8 \times 8$ eddies). In all the simulations 
presented in this figure we have employed $D=0.3$. } 
\label{fig3} 
\end{figure}

\begin{figure}
\centerline{\psfig{figure=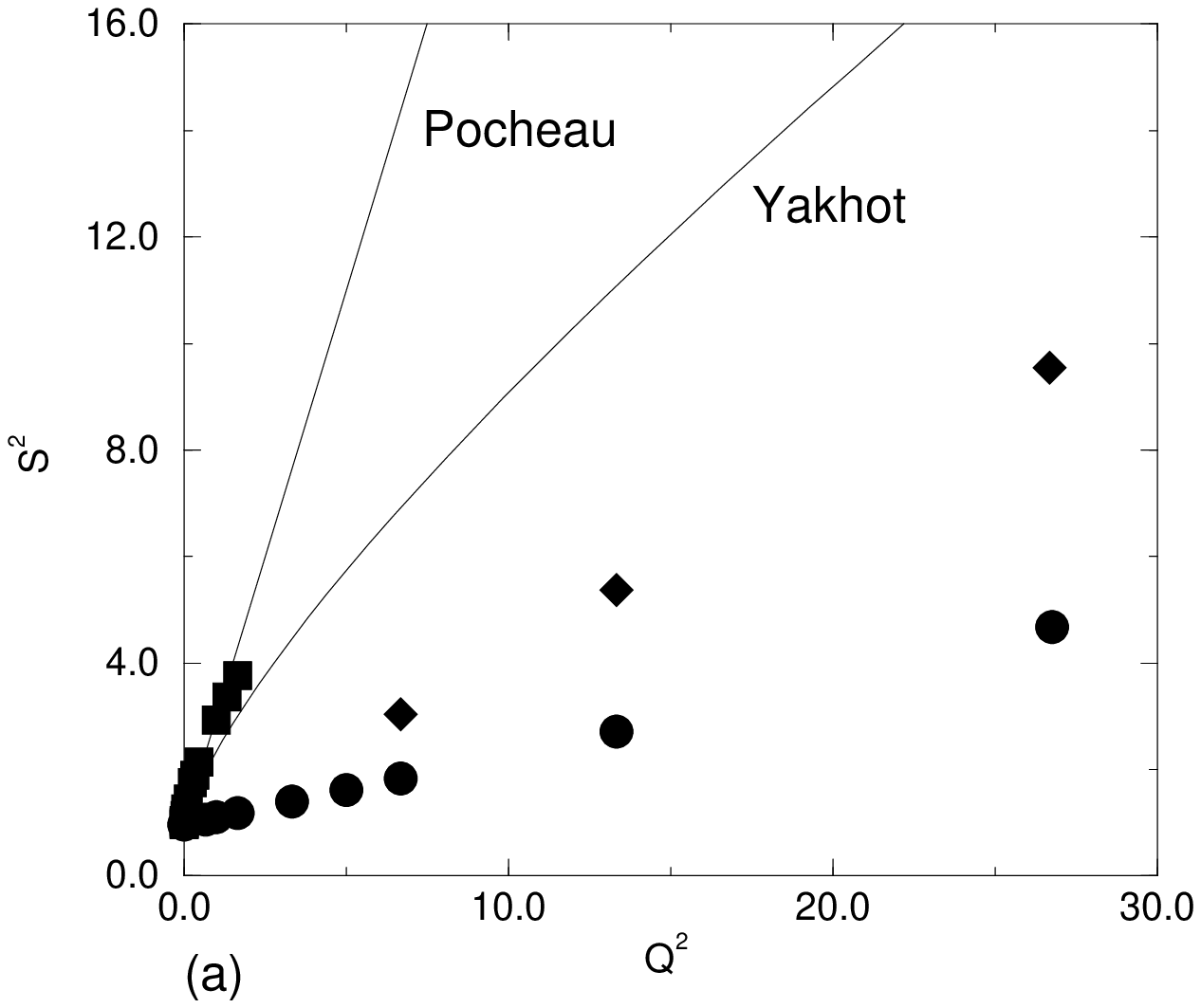,height=5cm,width=8cm}}
\centerline{\psfig{figure=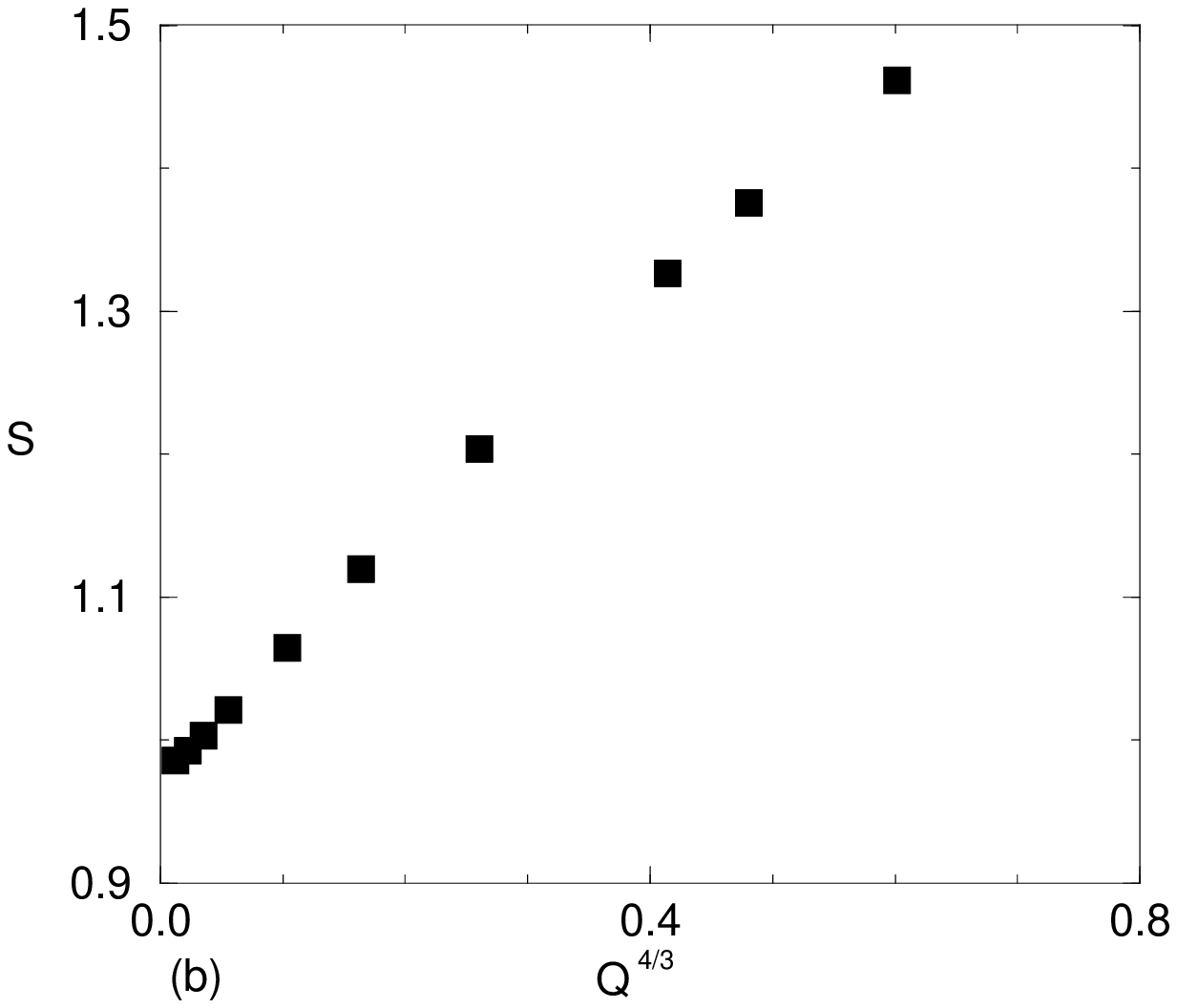,height=5cm,width=8cm}}
\centerline{\psfig{figure=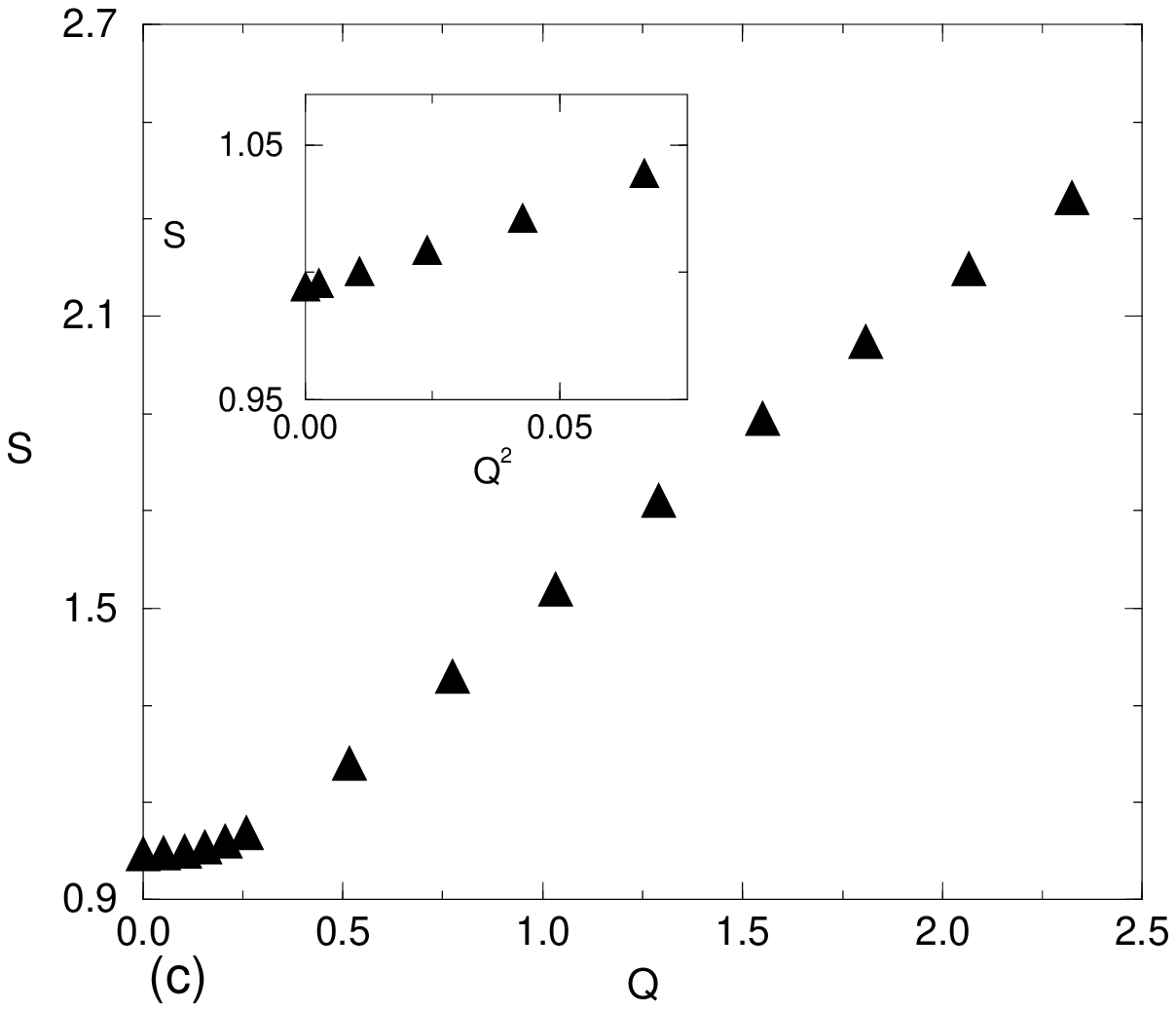,height=5cm,width=8cm}}
\caption{$S(Q)$ for three stirring modes: turbulent flows (circles, $t_0=0.1$;
romboids,
$t_0=3.0$), frozen stirring (squares), and periodic eddies (triangles). 
The same parameter values of 
Fig.~\protect\ref{fig3} have been employed. Continuous lines stand for the
theoretical predictions (see the text).
a): Turbulent and frozen flows. b): Frozen flow and c): Periodic flow.
}
\label{fig4}
\end{figure}

\begin{figure}
\centerline{\psfig{figure=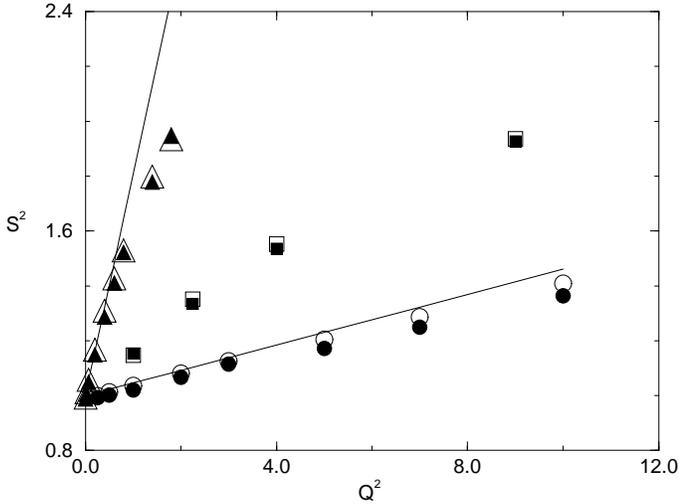,height=7cm}}
\caption{S vs $Q$ for DRZ regime, full symbols denote simulation
results of the front propagation while open symbols stand for simulation
of the pure effective diffusion
Eq.~\protect\ref{DRZ:eq}
(without reaction).
In all the simulations we have employed 
$D=2.0$.
Circles, squares and triangles correspond to  turbulent flow ($l_0=2.0$,
$t_0=0.1$), frozen stirring ($l_0=2.0$) 
 and periodic eddies respectively ($16 \times 16 $ eddies).
Continuous lines stand for theoretical predictions (see text).}
\label{fig5}
\end{figure}

\begin{figure}
\centerline{\psfig{figure=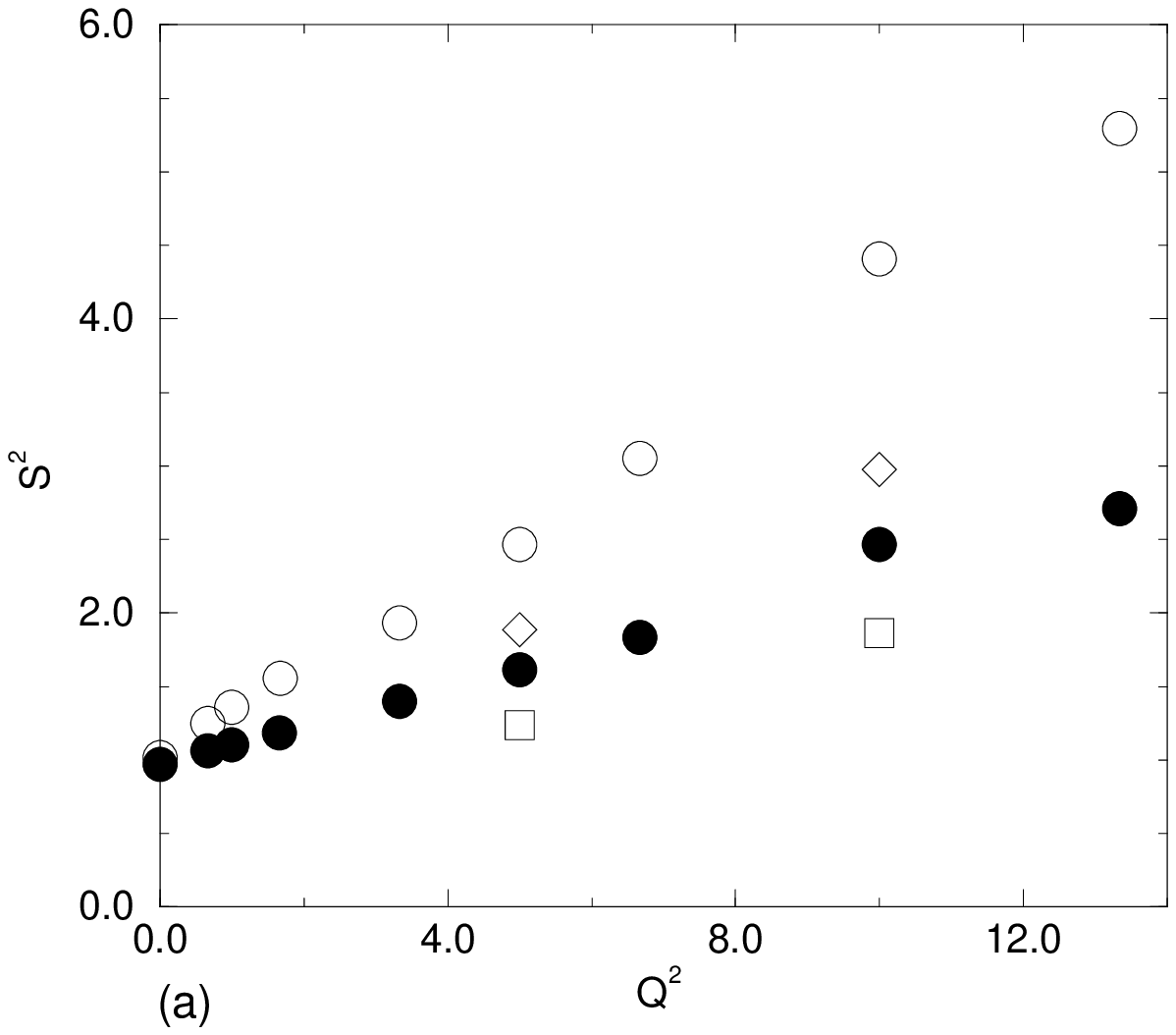,height=6cm,width=8cm}}
\centerline{\psfig{figure=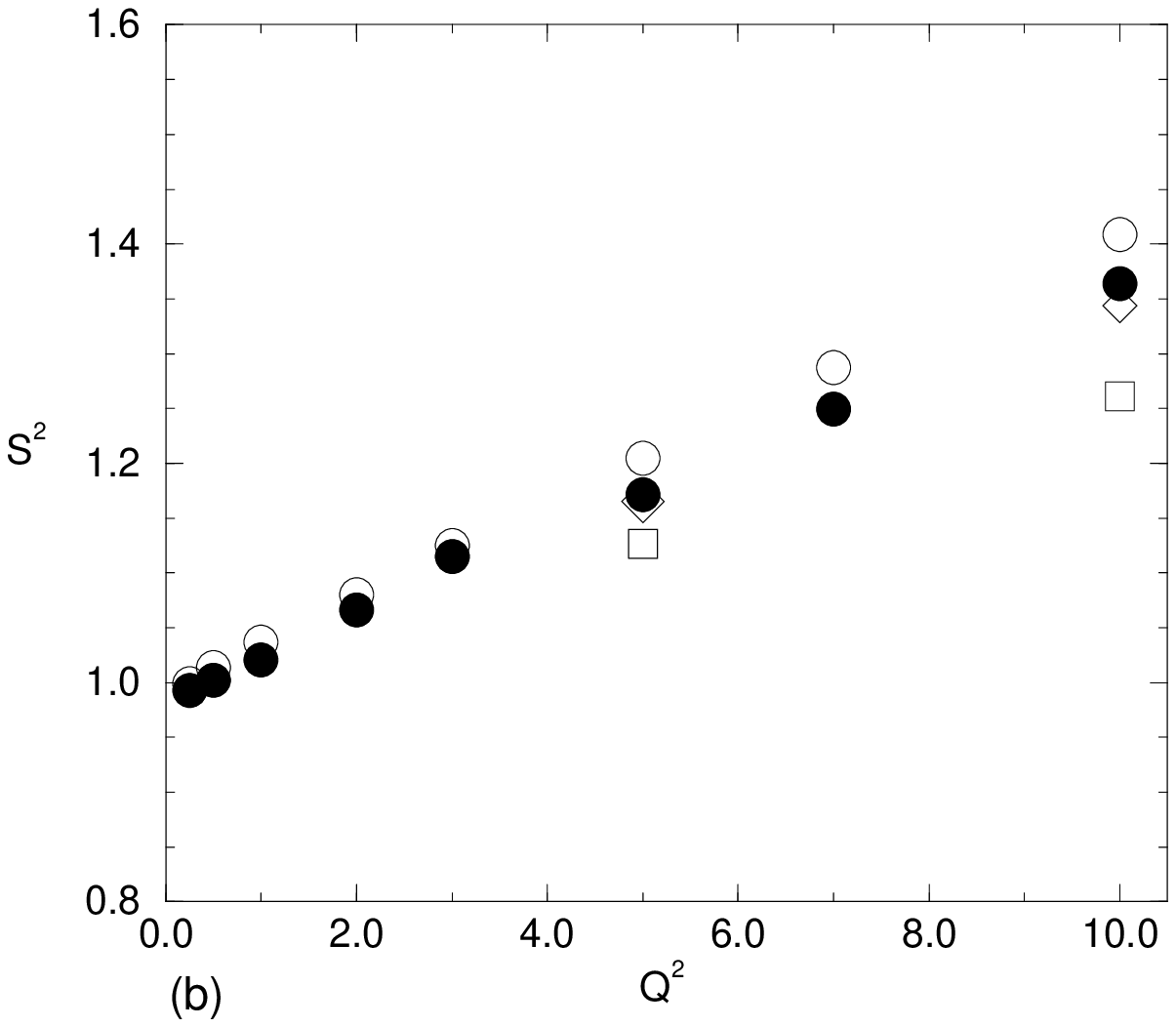,height=6cm,width=8cm}}
\caption{Comparison between K and  KO spectra.
Circles stand for the K spectrum while 
squares and romboids 
stand for the  KO spectrum with the same
$l_0$ and the same $k_0$ respectively (see Fig.~\protect\ref{fig1}).
Black circles are the same simulation data of 
Figs.~\protect\ref{fig4}a and 5, and white circles 
correspond to the simulated effective diffusion. 
(a) Thin front propagating mode.
(b) Distributed front propagating mode.
For KO spectrum we have employed $N=256$.} 
\label{fig6}
\end{figure}

In order to investigate the influence of the different spectra on the
enhancement of the front speed, we present in Fig.~\ref{fig6}
numerical results corresponding to the two regimes of front
propagation previously identified and the pair of turbulent spectra
introduced in Sec.~\ref{sec:lang}.  Respectively, Fig.~\ref{fig6}a
reproduces conditions of thin front propagation, whereas
Fig.~\ref{fig6}b those for distributed reaction fronts. Both in
Figs.~\ref{fig6}a and~\ref{fig6}b, circles stand for results obtained
with K spectrum, whereas squares and romboids correspond to the KO
distribution. Since both spectra can be compared either on the basis
of their integral length scale, $l_0$, or of their maximum, $k_0$, we
have chosen to separately consider both cases.  Specifically, squares
represent simulations from both spectra with the same value of $l_0$,
i. e. comparing cases a) and b) in Fig.~\ref{fig1}, and romboids
correspond to both spectra taken at equal values of $k_0$, i.e.
comparing cases a) and c) in Fig.~\ref{fig1}. On the other hand, black
circles in Fig.~\ref{fig6}a and black and white ones in
Fig.~\ref{fig6}b are respectively replotted from Figs.~\ref{fig4}a
and~\ref{fig5}.  Just for the sake of comparison, we have also
included in Fig.~\ref{fig6}a as white circles, results for $S^2$
obtained from the simulated values of the effective diffusion through
the use of expression (\ref{DRZ:eq}).
	
Let us start with the conditions of the HP regime
(Fig.~\ref{fig6}a). When both spectra are taken at equal $l_0$, we
find smaller values of the turbulent front velocity for KO than for K
spectrum. The situation is just the reverse one when both spectra are
taken with the same maximum. Both findings admit a rather direct
interpretation. From Fig.~\ref{fig1}a we conclude that when both
spectra have equal integral length scale, the KO distribution displays
its maximum at a smaller value of $k_0$. This means that in this last
case larger length scale modulations are going to be the most relevant
ones in bending the propagating interface.  This in turn means that
the front length is going to be less enhanced and, since we are
referring to the HP propagation mode, so will happen with the
effective front speed.  Contrarily when the two maxima coincide, what
makes the difference is the long tail in the KO distribution. Such
large wavenumbers, small spatial modulations, are going to be very
effective both in wrinkling more intensively the front interface but
even more, by interfering the front dynamics at scales comparable to
the front thickness. Both effects lead to larger values of the
turbulent front velocity. In particular the last mentioned feature is
evidenced when observing that results for the KO spectrum approach in
this case those that would correspond to a front subjected to the K
distribution and propagating under the enhanced diffusion mechanism
(white circles). 
A somewhat similar enhancement of the front velocity at moderate $Q$
has been found in a theoretical analysis \cite{Ronney92} when a thin 
front is slightly perturbed at scales smaller than $\delta_0$.

Let us turn now to the DRZ regime of Fig.~\ref{fig6}b. What we observe
in this situation is that no appreciable effect is found when both
maxima coincide in the respective spectra.  However, smaller values of
$S^2$ are observed for equal integral length scales. Analogously as
before, both results can be interpreted just by looking at
Fig.~\ref{fig1}. In the first case, one can imagine that the interface
thickness would be wide enough to comprise the most energetic
wavenumbers of both flows in such a way that the effective diffusion
would be similarly renormalized. Contrarily when prescribing an equal
value of $l_0$, those more energetic modes in the KO distribution are
shifted to smaller wavenumbers and in turn are going to be less
effective in enhancing the effective diffusion.

\section{Summary and Perspectives}
\label{sec:summ}

In summary, we have presented here a simple model of a chemical front
propagating in a turbulent media. Although artificial, the model
reproduces realistic scenarios of front propagation modes in chemical
experiments.  The observed dependences of the propagating front velocities
on the parameters and characteristics of the flow compares well with some
of the most well-known relations proposed in the literature. The
simplicity and versatility of the computer implementation of turbulent
flows and the simple model used to describe the front propagation in these
media open new perspectives in the study of more complicated situations in
turbulent fluids. 

\begin{acknowledgements}
We acknowledge financial support by Comision Interministerial de Ciencia y
Tecnologia, (Projects, PB93-0769, PB93-0759) and Centre de Supercomputaci\'o
de Catalunya, Comissionat per Universitats i Recerca de la Generalitat
de Catalunya. A.C.M. also acknowledges partial support from  the CONICYT
(Uruguay) and the Programa Mutis (ICI, Spain). We also acknowledge P.D. 
Ronney for fruitful comments.
\end{acknowledgements}


\begin{references}

\bibitem{pelce} P. Pelc\'e, {\it Dynamics of Curved Fronts},
Persp. in Physics (Academic Press, 1988).

\bibitem{murray} J.D. Murray, {\it Mathematical Biology}
(Springer, New York, 1989). 

\bibitem{cross}  M.C. Cross and  P.C. Hohenberg, ``Pattern formation
outside equilibrium,'' Rev. Mod. Phys. {\bf 65}, 851 (1993).

\bibitem{Williams85}
F.A. Williams, {\it Combustion Theory}, 2nd ed.
(Benjamin-Cummins, Menlo Park, California, 1985).

\bibitem{Pope87}S.B. Pope,
``Turbulent premixed flames,''
Annu. Rev. Fluid Mech. {\bf 19}, 237 (1987).

\bibitem{Ronney95a} P.D. Ronney,
``Some open issues  in premixed turbulent combustion,''
in Modeling in Combustion Science, J. Buckmaster and T. Takeno Eds,
Lectures notes in Physics, vol. 449 (Springer-Verlag, Berlin, 1995).

\bibitem{Kernstein88}
A.R. Kernstein, W.T. Ashurst, and F.A. Williams,
``Field equation for interface propagation in a unsteady homogeneous flow
field,'' Phys. Rev. A {\bf 37}, 2728 (1988).
For a recent review see also 
P.F. Embid, A.J. Majda, and P.E. Souganidis,
``Comparison of turbulent flame speeds from complete averaging
and the G-equation,'' Phys. Fluids {\bf 7}, 2052 (1995).

\bibitem{Kernstein92}
A.R. Kernstein and W.T. Ashurst,
``Propagation rate of growing interfaces in stirred fluids,''
Phys. Rev. Lett. {\bf 68}, 934 (1992).

\bibitem{Majda96}
S.P. Fedotov, ``Reaction front propagation in a turbulent flow,''
J. Phys. A: Math. Gen. {\bf 28}, L461 (1995),
A.J. Majda and P.E. Souganidis, ``Bounds on enhanced 
turbulent flame speeds for combustion with fractal velocity fields,''
J. Stat. Phys. {\bf 83},  933 (1996). 

\bibitem{Pocheau}
A. Pocheau, ``From turbulent flame fronts to their wrinkling process: 
scale invariance and similarity,''
C.R. Acad. Sci. Paris {\bf 319}, s II, 879 (1994),
A. Pocheau and D. Queiros-Cond\'e, ``Scale covariance 
of the wrinkling law of turbulent propagating interfaces,''
Phys. Rev. Lett. {\bf 76}, 3352 (1996).

\bibitem{Shy92} S.S. Shy, P.D. Ronney, S.G. Buckley, and V. Yakhot,
``Experimental simulation of premixed turbulent combustion using
aqueous autocatalytic reactions,'' {\it Proceedings of the 24th
Symposium (International) on Combustion} (Combustion Institute,
Pittsburgh, 1992).

\bibitem{Ronney95}
P.D. Ronney, B.D. Haslam, and N.O. Rhys,
``Front Propagation rates in randomly stirred media,''
Phys. Rev. Lett. {\bf 74}, 3804 (1995),
B.D. Haslam and P.D. Ronney,
``Fractal properties of propagating fronts in a strongly stirred fluid,''
Phys. Fluids {\bf7}, 1931 (1995).

\bibitem{Batchelor70}  G.K. Batchelor, {\it The Theory of Homogeneous
Turbulence} (Cambridge University, Cambridge, 1970).

\bibitem{McComb90and95}W.D. Mc Comb, ``Theory of turbulence,''
Rep. Prog. Phys. {\bf 58}, 1117 (1995); W.D. Mc Comb {\em The Physics 
of Fluid Turbulence} (Oxford University, Oxford, 1990).

\bibitem{Juneja94}A. Juneja, D.P. Lathrop, K.R. Sreenivasan, and
G. Stolovitzky, ``Synthetic turbulence,'' Phys. Rev. E {\bf 49},
5179 (1994);  M. Holzer and E.D. Siggia, ``Turbulent mixing of a passive
scalar,'' Phys. Fluids {\bf 6}, 1820 (1994).

\bibitem{marti} 
A. Careta, F. Sagu\'es, and J.M. Sancho,
``Generation of homogeneous isotropic turbulence with
well-defined spectra,'' Phys. Rev. E  {\bf 48}, 2279 (1993); A.C. Mart\'{\i},
J.M. Sancho, F. Sagu\'es, and A. Careta ``Langevin approach to generate 
synthetic turbulent flows,'' Phys. Fluids, {\bf 9}, 1078 (1997).

\bibitem{Yakhot88}V. Yakhot,
``Propagation velocity of premixed turbulent flames,''
Combust. Sci. Tech. {\bf 60}, 191 (1988).

\bibitem{Haworth92} 
D.C. Haworth and T.J. Poinsot, ``Numerical simulations of Lewis 
number effects in turbulent premixed flames,''
J. Fluid Mech. {\bf244}, 405 (1992).

\bibitem{Kraichnan70}R.H. Kraichnan,``Diffusion by
a random velocity field,'' Phys. Fluids {\bf 13}, 22 (1970).

\bibitem{K-O} T.V. von K\'arman,``Progress in the theoretical
theory of turbulence,'' Proc. Natl. Acad. Sci.
U.S.A. {\bf 34}, 530 (1948); A.M. Obukhov, ``On the distribution
of energy in the spectrum of turbulent flow,'' C.R. Acad. Sci.
USSR  {\bf 32}, 19 (1941).

\bibitem{Moffat} H.K.  Moffat,
``Transport effects associated with turbulence with particular
attention to the influence of the helicity,'' 
Rep. Prog. Phys. {\bf 46}, 621 (1993).

\bibitem{Ronney92}P.D. Ronney and V. Yakhot,
``Flame broadening effects on premixed turbulent flame speeds,''
Combust. Sci. and Tech. {\bf 86}, 31 (1992).

\end{references}
\end{document}